\documentclass[10pt, conference, letterpaper]{IEEEtran}
\ifCLASSINFOpdf
\else
\usepackage[dvips]{graphicx}
\fi
\usepackage[cmex10]{amsmath}
\usepackage{upgreek}
\usepackage{booktabs}
\usepackage{epsfig}
\usepackage{latexsym}
\usepackage{multirow}
\usepackage{stfloats}
\usepackage{epstopdf}
\usepackage{color}  
\usepackage{tabularx} 
\usepackage{amssymb}
\usepackage{enumerate}
\graphicspath{{./Figures/}}
\usepackage{color}
\usepackage{bbm}
\usepackage{bm}
\usepackage{cite}
\usepackage{balance}
\usepackage{mathrsfs}
\usepackage{verbatim}
\usepackage{dsfont}
\usepackage{verbatim}
\usepackage{tikz}
\usepackage{algorithm}
\usepackage{algpseudocode}
\usepackage{diagbox}
\allowdisplaybreaks[4]
\usepackage[framemethod=tikz]{mdframed}
\usepackage{multicol}
\usepackage{environ}
\usepackage{tikz}

\usepackage{subfig}
\title{L-COIN: LLM-Assisted Counterfactual Inference for Game-Theoretic Distributed Computation Offloading in Sub-THz LEO Satellite Networks
}

\author{
  \IEEEauthorblockN{Jinhao Yi, Weijun Gao, and Chong Han}  
  \IEEEauthorblockA{Shanghai Jiao Tong University, Shanghai, China.}
  \IEEEauthorblockA{E-mail: \{jinhao.yi, gaoweijun, chong.han\}@sjtu.edu.cn}  
}

\begin{document}

\maketitle

\begin{abstract} 

As Space-Based Information Networks (SBINs) evolve toward high-capacity, intelligence-centric paradigms, integrating sub-Terahertz (sub-THz) communication into Low Earth Orbit (LEO) satellite constellations has emerged as a critical enabler for ultra-broadband and resilient global connectivity. By exploiting the ultra-wide bandwidth of sub-THz links to reduce transmission delays, resource-constrained ground devices can seamlessly offload compute-intensive tasks to LEO edge servers. However, satellite motion, short visibility windows, and limited onboard resources make offloading decisions highly time-varying. Existing distributed offloading schemes typically require repeated inter-device state exchange and poorly adapt to time-varying LEO topology or traffic conditions. To address these limitations,  a decentralized game-theoretic offloading framework empowered by large language models (LLMs) and counterfactual inference is proposed in this paper.
First, a realistic offloading system is established by integrating time-varying 3D-Walker topology. Second, a game-theoretic scheme using counterfactual inference is introduced to deduce unobserved states from local histories, eliminating global information reliance. Finally, an LLM-empowered semantic fusion algorithm is integrated into the counterfactual inference to enhance adaptability through zero-shot reasoning and self-reflection. Numerical results show that L-COIN reduces offloading cost by 10.9\% to 27.7\% relative to state-of-the-art baselines.
\end{abstract}


\section{Introduction}

\IEEEPARstart{A}{s} Space-Based Information Networks (SBINs) continuously evolve toward high-capacity and intelligence-centric paradigms, the demand for ultra-broadband and resilient global connectivity has experienced unprecedented growth~\cite{zhou2023aerospace,yang20196g,xiao2024space}. To fulfill these escalating demands, integrating sub-Terahertz (sub-THz) communication into Low Earth Orbit (LEO) satellite-ground networks has emerged as a pivotal technology for 6G and beyond systems~\cite{luo2024leo,zheng2023sdn}. Operating at $0.1\sim0.3~\textrm{THz}$, sub-THz communication delivers unprecedented data rates and massive channel capacity~\cite{han2021terahertz}, significantly reducing task-upload and result-delivery times. Driven by these high-speed connectivity advancements, modern LEO satellites are transcending their traditional roles from mere communication relays to ubiquitous aerial servers for edge computing. Concurrently, as ground devices (e.g., IoT nodes) increasingly generate compute-intensive and latency-sensitive tasks~\cite{centenaro2021survey,abkenar2022survey}, their constrained local resources have become a severe bottleneck~\cite{tang2021computation}. By offloading these intensive workloads directly to the LEO edge, ground devices can effectively overcome local hardware limitations, thereby extending proximate computing capabilities on a global scale~\cite{kawamoto2023traffic,zhang2023long,li2024skycastle}.

However, the effective implementation of computation offloading in LEO-ground networks faces several severe challenges. First, unlike static terrestrial scenarios, LEO satellites orbit at exceptionally high speeds, leading to highly dynamic network topologies and intermittent communication links with strictly restricted visibility windows~\cite{cheng2019space}. Second, the inherent computing bottlenecks of satellites, coupled with the stochastic arrival of intensive workloads, inevitably lead to acute resource contention and extended queuing delays~\cite{zhang2023energy,chen2023energy}. Third, given the massive number of ground devices and the high-dimensional decision space~\cite{huang2022distributed}, implementing traditional centralized paradigms~\cite{shi2020priority,apostolopoulos2020risk} as they incur unacceptable communication overhead and severe transmission delays to gather global state information. 

To address these challenges, recent works have mainly focused on distributed paradigms that shift decision-making from a central controller to individual ground devices. In~\cite{chen2025game, zhou2022stackelberg}, game-theoretic schemes were developed to maximize the utility of edge servers. The authors in~\cite{zhou2024mobility,jiang2022joint} adopt alternating direction method of multipliers (ADMM) and Lyapunov optimization to reduce the offloading cost. Meanwhile, other studies~\cite{huang2026joint,xu2025edge,li2024computation} focused on multi-agent deep reinforcement learning~(MADRL) frameworks, where each device learns a local offloading policy, thereby avoiding real-time global state exchange during the execution phase.
However, these schemes still suffer from several critical limitations. First, most rely on simplified or static network topology assumptions, failing to capture the highly dynamic and time-varying nature of real satellite networks. Second, conventional optimization-based schemes still require repeated inter-device state exchange. Third, MADRL-based schemes demand massive training datasets that are exceptionally difficult to acquire. Moreover, most of the aforementioned schemes struggle to adapt to time-varying LEO topology or traffic conditions deviating from their initial design

Motivated by these limitations, in this paper, we propose a decentralized game-theoretic computation offloading framework empowered by \underline{\textbf{L}}LM-assisted \underline{\textbf{Co}}unterfactual \underline{\textbf{I}}nference in LEO-ground \underline{\textbf{N}}etworks~(\textbf{L-COIN}). Specifically, we first establish a realistic offloading system integrating time-varying 3D-Walker constellation dynamics to capture the dynamics in LEO-ground networks. 
Second, we introduce a counterfactual inference-based decentralized game-theoretic offloading scheme, allowing ground devices to deduce unobserved states solely from local histories, thereby eliminating the reliance on global information. Furthermore, 
we develop a Large Language Model (LLM)-empowered semantic fusion and self-reflection algorithm that is seamlessly integrated into the counterfactual inference scheme. By leveraging the zero-shot reasoning and semantic reflection capabilities of LLMs, this training-free algorithm adaptively enhances generalization of counterfactual inference, ensuring autonomous convergence to a Nash Equilibrium without massive datasets. Specifically, the contributions of this paper are summarized as follows.

\begin{figure*}[t]
\centerline{\includegraphics[width=0.7\textwidth]{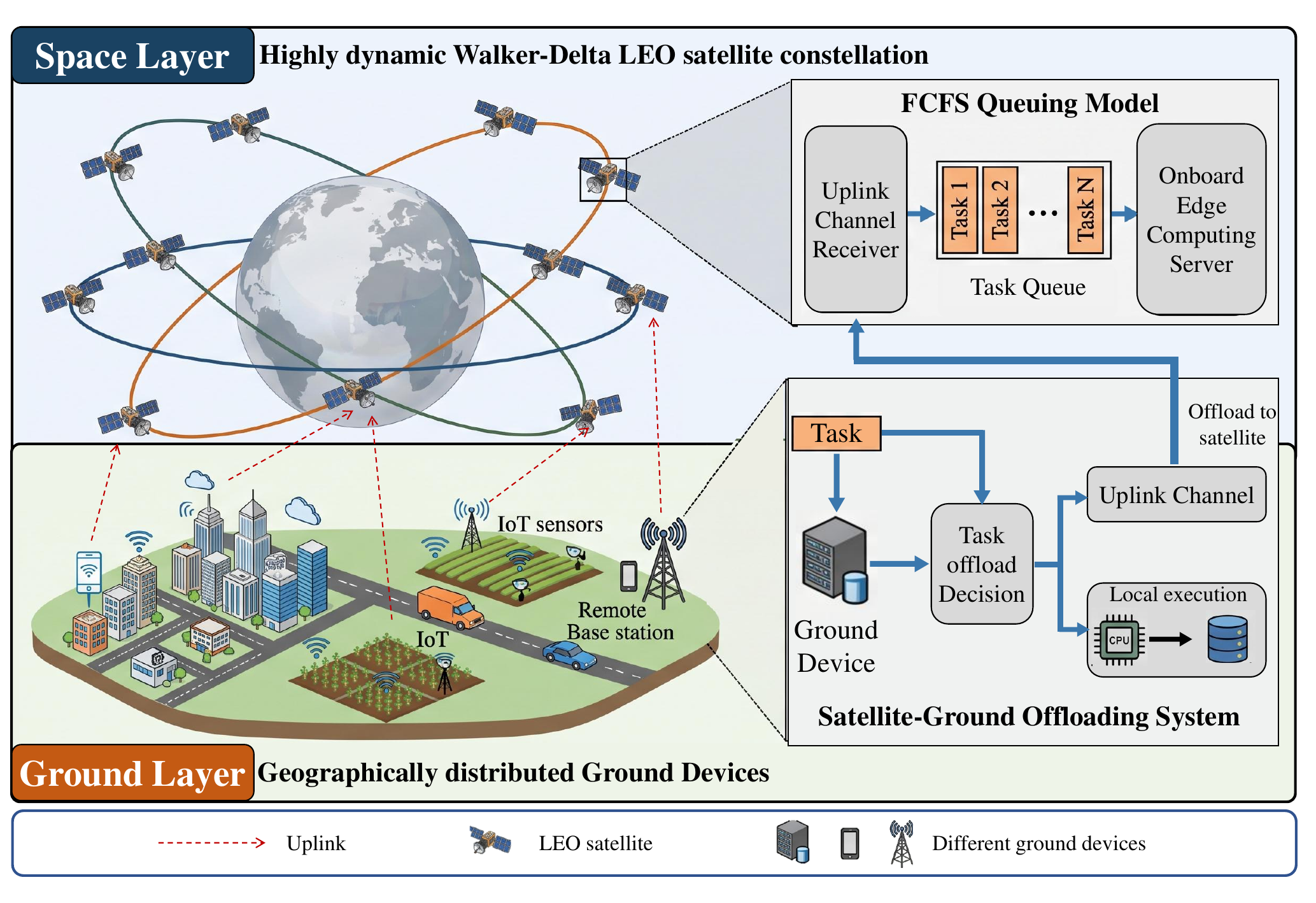}}
\captionsetup{font={footnotesize}}
\caption{System model of the two-tier LEO-ground edge computing.}
\label{fig:system_model}
\end{figure*}

\begin{itemize}
\item \textbf{We propose a realistic computation offloading system model that captures the spatiotemporal dynamics of LEO-ground networks.} Unlike many existing works that rely on simplified or static topology assumptions, our model integrates time-varying 3D-Walker constellation dynamics and a First-Come-First-Serve~(FCFS) queuing mechanism. This integration enables accurate characterization of the spatiotemporal dynamics in real-world LEO satellite systems.

\item \textbf{We formulate a decentralized exact potential game offloading framework empowered by counterfactual inference.} To overcome the reliance on impractical global information, we introduce a dual-source counterfactual inference mechanism into the decentralized game-theoretic offloading framework. This mechanism allows each ground device to infer the unobserved states of other competing devices from its own local historical observations, thereby enabling fully decentralized decision-making. Consequently, each device can independently converge to its optimal offloading strategy without requiring real-time state exchange.

\item \textbf{We introduce a training-free semantic fusion algorithm driven by LLMs into the counterfactual inference mechanism.} To overcome the data dependency and poor adaptation of existing schemes, we integrate LLMs into the counterfactual inference scheme. By exploiting the inherent zero-shot reasoning and semantic reflection capabilities of LLMs, this algorithm adaptively fuses semantic knowledge to enhance inference accuracy across diverse and previously unseen network configurations, allowing the game to autonomously converge to a Nash Equilibrium without requiring massive pre-collected training data or iterative retraining.
\end{itemize}

The remainder of this paper is structured as follows. In Sec.~\ref{System_Model}, we present the system model for LEO-ground edge computing. In Sec.~\ref{Algorithm}, we formulate the decentralized offloading problem and propose the L-COIN framework. Numerical results are demonstrated in Sec.~\ref{Numerical}, and Sec.~\ref{Conclusion} concludes this paper.

\section{System Model}
\label{System_Model}
As illustrated in Fig.~\ref{fig:system_model}, we consider a two-tier LEO-ground offloading system comprising a ground layer of geographically distributed ground devices (GDs) and a space layer featuring a highly dynamic Walker-Delta LEO satellite constellation~\cite{wei2020efficient}. The GDs generate tasks that can either be processed locally or offloaded to visible satellites via uplink channels. Additionally, to manage concurrent offloading requests, each satellite maintains a task queue. The detailed description of the system model is presented as follows.

\subsection{Satellite Network Model}
In our scenario, We model the space layer as a Walker-Delta LEO satellite constellation orbiting at an altitude $h$. This constellation is uniquely parameterized by a standard four-tuple, i.e., the number of orbital planes~$P$, the number of satellites per plane $T$, the phasing factor $F \in \{0, 1, \ldots, P-1\}$, and the orbital inclination $\theta$. Accordingly, the right ascension of the ascending node (RAAN) $\Omega^m_i$ and the initial mean anomaly $u^m_i$ of the $i$-th satellite in the $m$-th plane are governed by
\begin{equation}
\left\{\begin{array}{l}
\Omega^m_i=\Omega_{0}+(m-1) \frac{2 \pi}{P}, \\
u^m_i=u_{0}+(m-1) F \frac{2 \pi}{T}+(i-1) P \frac{2 \pi}{T},
\end{array}\right.
\label{eq:orbit_params}
\end{equation}
where $\Omega_{0}$ and $u_{0}$ denote the reference RAAN and mean anomaly, respectively. Based on these orbital mechanics, the 3D position $\mathbf{p}_s(t)$ of the $i$-th satellite in the $m$-th orbital plane at time $t$ in the Earth-Centered Inertial (ECI) coordinate system is formulated as
\begin{equation}
\mathbf{p}_s(t) = R \cdot \begin{bmatrix}
\cos u^m_i(t) \cos \Omega^m_i - \sin u^m_i(t) \sin \Omega^m_i \cos \theta \\
\cos u^m_i(t) \sin \Omega^m_i + \sin u^m_i(t) \cos \Omega^m_i \cos \theta \\
\sin u^m_i(t) \sin \theta
\end{bmatrix},
\label{eq:satellite_position}
\end{equation}
where $R = R_e + h$ represents the orbital radius ($R_e$ being the Earth's radius). $u^m_i(t) = u^m_i + \omega_s t$ captures the time-varying mean anomaly, where $\omega_s = \sqrt{\mu / R^3}$ is the orbital angular velocity, and $\mu$ is the Earth's gravitational constant.

To capture the fast mobility of LEO satellites, we adopt a discrete-time snapshot model. The continuous operational timeline is divided into sequential time slots $t \in \mathcal{T} = \{1, 2, \ldots, N\}$ with a sufficiently small duration $\tau$. Within each time slot, the LEO network is considered quasi-static, while the physical state evolves strictly across consecutive slots.

\subsection{Sub-THz Communication Model}
To enable ultra-high-speed transmission, we adopt sub-THz communications for the LEO-ground links. While sub-THz provides an ultra-wide bandwidth $B$, its channel capacity is constrained by path loss, molecular absorption, and beam misalignment~\cite{Masihi2025Terahertz}. Therefore, we establish a comprehensive sub-THz communication model that captures these physical losses, multi-user interference, and dynamic visibility constraints.

First, the composite channel gain $h_{n,s}(t)$ between ground device $n$ and satellite $s$ is evaluated as
\begin{equation}
    h_{n,s}(t) = G_{tx} G_{rx} \left(\frac{c}{4 \pi f_c d_{n,s}(t)}\right)^2 \frac{h_{pe}(t)}{\Psi_{n,s}(t)},
    \label{eq:channel_gain}
\end{equation}
where $c$ is the speed of light, $f_c$ is the sub-THz carrier frequency, and $G_{tx}, G_{rx}$ are the antenna gains.  The term $\Psi(f, \mathbf{r}_i, \mathbf{r}_m)$ represents the cumulative multi-species molecular absorption along the propagation path, which is given as 
\begin{equation}
\Psi(f, \mathbf{r}_i, \mathbf{r}_n) = \exp\left( \int_{0}^{\|\mathbf{r}_i(t) - \mathbf{r}_n\|} \sum_{q} \kappa_a^{(q)}(f, \ell) d\ell \right),
\end{equation}
where $\kappa_a(f,r_{atm}) = \sum_i\kappa_a^i(f,r_{atm})$ is the total summed molecular absorption coefficient along the propagation path through the atmosphere. In addition, $h_{pe}(\alpha(t))$ represents the pointing loss factor due to beam misalignment, which can be expressed as~\cite{Masihi2025Terahertz}
\begin{equation}
h_{pe}(\alpha(t)) = \exp\left(-2 \frac{(\|r_i(t)-r_n\| \tan \alpha(t))^2}{w_{zeq}^2}\right)
\end{equation}
where $\alpha(t)$ is the instantaneous pointing error angle, and $w_{zeq}^2$ is the equivalent beam waist given in~\cite{Masihi2025Terahertz}.

Second, in our multi-user scenario, concurrent offloading to the same satellite inherently induces co-channel interference. Let $\mathcal{N}_s(t)$ denote the subset of ground devices opting to offload their tasks to satellite $s$ during time slot $t$. For each device $n \in \mathcal{N}_s(t)$, the signal-to-interference-plus-noise ratio (SINR) is formulated as
\begin{equation}
    \gamma_{n,s}(t) = \frac{P_n h_{n,s}(t)}{\sum_{n' \in \mathcal{N}_s(t) \setminus \{n\}} P_{n'} h_{n',s}(t) + \sigma^2},
    \label{eq:sinr}
\end{equation}
where $P_n$ is the transmission power of device $n$, and $\sigma^2$ is the Gaussian white noise. Hence, the achievable uplink data rate for device $n$ transmitting to LEO satellite $s$ is given as
\begin{equation}
    R_{n,s}(t) = B \log_2 \big( 1 + \gamma_{n,s}(t) \big).
    \label{eq:data_rate}
\end{equation}
Therefore, for a computation task generated by device $n$ with an input data size $D_n$, the time $T_{n,s}^{\text{upload}}$ required to upload this task to satellite $s$ is given as
\begin{equation}
    T_{n,s}^{\text{upload}} = \frac{D_n}{R_{n,s}(t)} + \frac{d_{n,s}(t)}{c},
\end{equation}
where the two terms denote the transmission delay and the propagation delay, respectively. 

Furthermore, to maintain a reliable sub-THz communication link, the target satellite $s$ must remain within the visible range of device $n$ throughout $T_{n,s}^{\text{upload}}$. Let $\mathbf{p}_n$ and $\mathbf{p}_s(t)$ denote the 3D coordinates of ground device $n$ and satellite $s$, respectively. The elevation angle $\beta_{n,s}(t)$ is given as
\begin{equation}
\beta_{n,s}(t) = \arcsin \left( \frac{\mathbf{p}_n \cdot \left(\mathbf{p}_s(t) - \mathbf{p}_n\right)}{\|\mathbf{p}_n\| \|\mathbf{p}_s(t) - \mathbf{p}_n\|} \right).
\label{eq:elevation_angle}
\end{equation}
Hence, to avoid severe atmospheric blockages and ensure reliable communication, the elevation angle must exceed a predefined minimum threshold $\theta_{\min}$ from the start of the transmission $t_n^{\text{start}}$ until its completion, which is formulated as
\begin{equation}
    \min \left[ \beta_{n,s}(t_n^{\text{start}}), \beta_{n,s}(t_n^{\text{start}} + T_{n,s}^{\text{upload}}) \right] \ge \theta_{\min}.
\end{equation}

\subsection{Queuing and Computation Model}
Due to constrained onboard computing capacity, LEO satellites cannot process massive offloaded tasks simultaneously. Thus, we explicitly model this execution dynamics using the FCFS queuing model.
Specifically, for each device $n \in \mathcal{N}_s(t)$, its task arrives at satellite $s$ at timestamp $t_{n,s}^{\text{arrive}} = t_n^{\text{start}} + T_{n,s}^{\text{upload}}$. We sort the arrival timestamps of all concurrent tasks in $\mathcal{N}_s(t)$ in ascending order:
\begin{equation}
    t_{\pi(1),s}^{\text{arrive}} \le t_{\pi(2),s}^{\text{arrive}} \le \dots \le t_{\pi(|\mathcal{N}_s(t)|),s}^{\text{arrive}}.
    \label{eq:arrival_order}
\end{equation}
where $\pi(k)$ denotes the device whose task is the $k$-th to arrive at satellite $s$. 
Under the FCFS model, the $k$-th task must wait for the completion of the satellite's pre-existing local workloads, denoted by a baseline delay $T_s^{\text{local}}$, and all $k-1$ preceding offloaded tasks. Thus, its queuing waiting delay $T_{\pi(k),s}^{\text{wait}}$ is formulated as
\begin{equation}
    T_{\pi(k),s}^{\text{wait}} = T_s^{\text{local}} + \sum_{j=1}^{k-1} T_{\pi(j),s}^{\text{comp}} = T_s^{\text{local}} + \frac{\sum_{j=1}^{k-1} X_{\pi(j)}}{f_s},
    \label{eq:wait_time}
\end{equation}
where $X_{\pi(j)}$ is the required CPU cycles for task $\pi(j)$, and $f_s$ is the computing capacity of satellite $s$.
Consequently, for any device $n \in \mathcal{N}_s(t)$, the total offloading delay $T_{n,s}^{\text{total}}$ experienced by device $n$ is accumulated as
\begin{equation}
    T_{n,s}^{\text{total}} = T_{n,s}^{\text{trans}} + T_{n,s}^{\text{prop}} + T_{n,s}^{\text{wait}} + T_{n,s}^{\text{comp}}.
    \label{eq:total_delay}
\end{equation}

\subsection{Problem Formulation}
Based on the established physical models, the objective of the computation offloading
problem is to jointly optimizes the execution delay, energy consumption, and satellite load balancing for all ground devices. Let $\mathbf{a}(t) = \{a_1(t), a_2(t), \dots, a_N(t)\}$ denote the joint strategy profile of all ground devices at time slot $t$, where $a_n(t) \in \{0\} \cup \mathcal{S}_n^{\text{vis}}(t)$, where $a_n(t) = 0$ indicates local task execution, and $a_n(t) = s \in \mathcal{S}_n^{\text{vis}}(t)$ designates offloading to a currently visible LEO satellite $s$. 
Therefore, the comprehensive cost function $C_n(a_n, \mathbf{a}_{-n})$ for device $n$ is formulated as
\begin{equation}
    C_n(a_n, \mathbf{a}_{-n}) = \alpha T_n^{\text{total}}(a_n) + \beta E_n^{\text{total}}(a_n) + \gamma L_n(a_n),
    \label{eq:cost_function}
\end{equation}
where $\alpha$, $\beta$, and $\gamma$ are non-negative weighting parameters, and $\mathbf{a}_{-n}$ denotes the concurrent decisions of all other devices. The total time delay $T_n^{\text{total}}(a_n)$ is given by \eqref{eq:total_delay} if $a_n \neq 0$, and evaluates to $T_n^{\text{wait, loc}} + X_n/f_n^{\text{loc}}$ if $a_n = 0$, where $T_n^{\text{wait, loc}}$ and $f_n^{\text{loc}}$ denote the local queuing delay and CPU capacity. The energy consumption $E_n^{\text{total}}(a_n)$ comprises the transmission energy and the computing energy, which is evaluated as~\cite{wu2022joint}
\begin{equation}
E_n^{total}(a_n)=\kappa X_n + \mathbb{I}(a_n\neq 0)P_n T_{n,s}^\text{trans},
\end{equation}
where $\kappa$ is the energy consumption per CPU cycle. $\mathbb{I}(\cdot)$ is an indicator function that equals 1 if the condition is true and 0 otherwise. Similarly, the load penalty $L_n(a_n)$, designed to mitigate satellite congestion, is defined as
\begin{equation}
L_n(a_n)=\mathbb{I}(a_n\neq 0)\frac{|\mathcal{N}_s(t)|}{Q_s^{max}},
\end{equation}
where $Q_s^{\max}$ is the maximum task queue capacity.

Based on this comprehensive metric, the overarching objective is to minimize the total system cost of all devices, which can be mathematically formulated as
\begin{align}
    (\mathcal{P}1): \quad \min_{\mathbf{a}} \quad & \sum_{n \in \mathcal{N}} C_n(a_n, \mathbf{a}_{-n}) \label{eq:objective} \\
    \text{s.t.} \quad 
    & a_n(t) \in \{0\} \cup \mathcal{S}_n^{\text{vis}}(t), \quad \forall n \in \mathcal{N}, \tag{\ref{eq:objective}a} \\
    & \min \left[ \beta_{n,s}(t_n^{\text{start}}), \beta_{n,s}(t_n^{\text{start}} + T_{n,s}^{\text{upload}}) \right] \ge \theta_{\min}, \nonumber \\
    & \quad \forall n \in \mathcal{N}, \text{ if } a_n = s \in \mathcal{S}_n^{\text{vis}}(t). \tag{\ref{eq:objective}b}
\end{align}
Here, constraint (\ref{eq:objective}a) strictly bounds the decision space to visible satellites, while constraint (\ref{eq:objective}b) guarantees spatial link reliability throughout the entire transmission window.

Since the cost $C_n(a_n, \mathbf{a}_{-n})$ is tightly coupled with the decisions of other devices $\mathbf{a}_{-n}$ via co-channel interference and FCFS queues, solving $(\mathcal{P}1)$ centrally incurs prohibitive signaling overhead and is practically infeasible. To address this challenge, we propose a decentralized exact potential game
offloading framework empowered by counterfactual
inference in the following section.

\begin{figure*}[t]
\centerline{\includegraphics[width=0.7\textwidth]{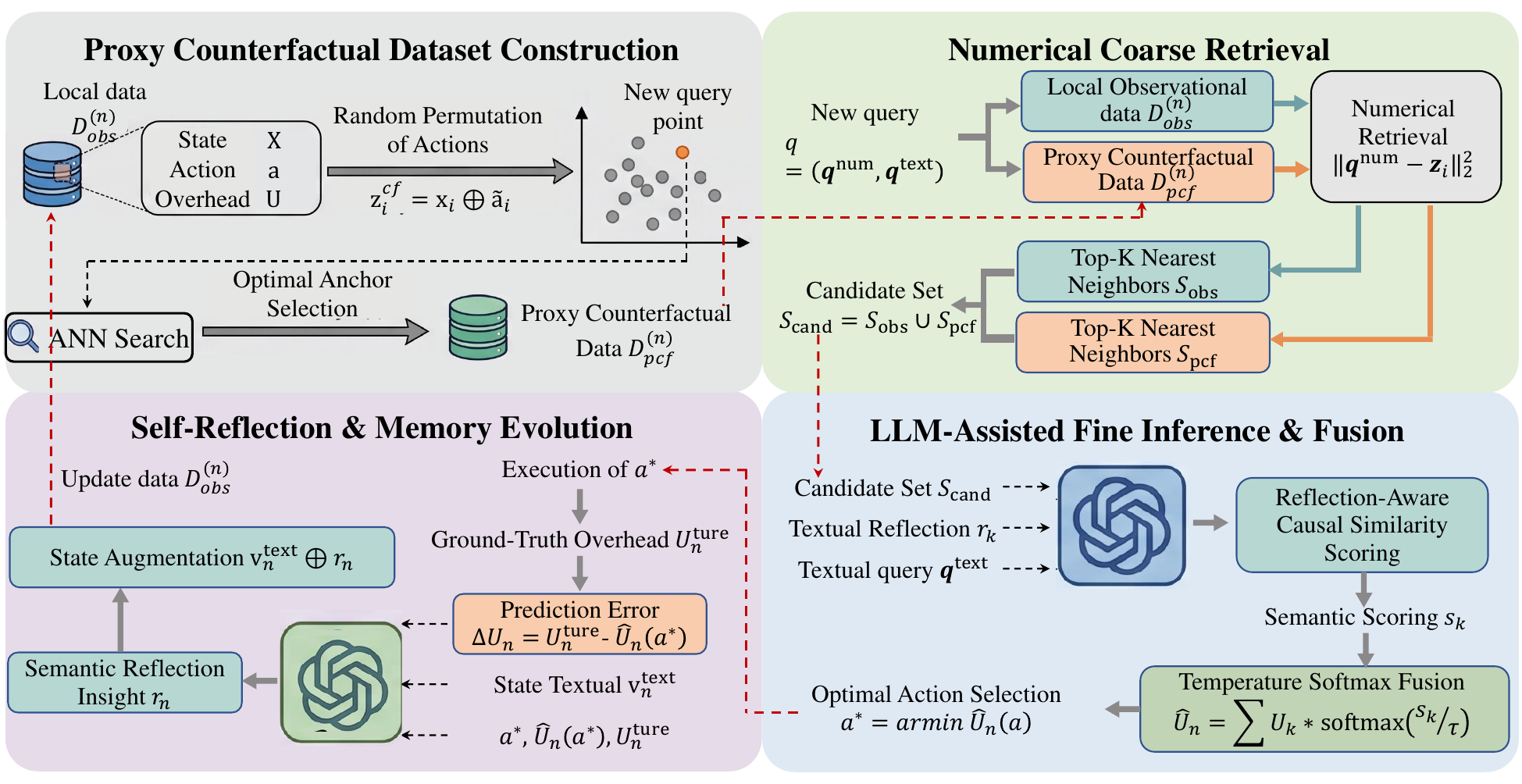}}
\captionsetup{font={footnotesize}}
\caption{Overview of L-COIN framework.}
\label{fig:algorithm}
\end{figure*}

\section{Decentralized Game-Theoretic Offloading via LLM-Assisted Counterfactual Inference}
\label{Algorithm}
In this section, we propose a fully decentralized game-theoretic framework driven by LLM-assisted counterfactual inference to address the coupled optimization problem $(\mathcal{P}1)$. Unlike existing decentralized schemes~\cite{chen2025game, zhou2022stackelberg,huang2026joint,xu2025edge,li2024computation} that require impractical global state information, we introduce a novel counterfactual inference mechanism, integrated with a training-free LLM-empowered semantic algorithm. This enables ground devices to evaluate unobserved strategies with only local histories, while simultaneously ensuring robust generalization even in unseen network environments.

\subsection{Decentralized exact potential game
offloading framework}
Motivated by~\cite{chen2025game}, we formulate the computation offloading problem as an exact potential game denoted by $\mathcal{G} = \langle \mathcal{N}, \{\mathcal{A}_n\}_{n \in \mathcal{N}}, \{C_n\}_{n \in \mathcal{N}} \rangle$~\cite{monderer1996potential}. In this framework, the players correspond to the set of all ground devices $\mathcal{N}$. For each device $n \in \mathcal{N}$, the available action space $\mathcal{A}_n = \{0\} \cup \mathcal{S}_n^{\text{vis}}(t)$ consists of executing the task locally or offloading it to one of the currently visible LEO satellites. $C_n(a_n, \mathbf{a}_{-n})$ represents the individual comprehensive cost of device $n$. To drive the system toward a globally optimal state, the global potential function $\Phi(\mathbf{a})$ is defined as  the optimization objective in $(\mathcal{P}1)$, which is given as
\begin{equation}
    \Phi(\mathbf{a}) = \sum_{n \in \mathcal{N}} C_n(a_n, \mathbf{a}_{-n}).
\end{equation}
However, optimizing this global potential function directly requires centralized coordination. To decouple the global objective and achieve fully distributed optimization, an auxiliary marginal overhead function $U_n(a_n, \mathbf{a}_{-n})$ is introduced for each device $n$, given as~\cite{chen2025game,blume1993statistical}
\begin{equation}
\begin{aligned}
    U_n(a_n, \mathbf{a}_{-n}) &= C_n(a_n, \mathbf{a}_{-n}) \\
    &+ \sum_{m \neq n} \bigl( C_m(a_m, \mathbf{a}_{-m}) - C_m(a_m, \mathbf{a}_{-m \backslash n}) \bigr),
\end{aligned}
\label{utility}
\end{equation}
where $C_m(a_m, \mathbf{a}_{-m \backslash n})$ represents the hypothetical cost of device $m$ assuming device $n$ does not exist. Consequently, the first term in~\eqref{utility} is the individual cost of device~$n$ itself, and the second term quantifies the negative externalities that device $n$ imposes on other devices at the current strategy~$\mathbf{a}$.
By constructing the individual overhead $U_n$ in this manner, it strictly preserves the exact potential game property, given as~\cite{chen2025game}
\begin{equation}
\begin{aligned}
    &U_n(a_n, \mathbf{a}_{-n}) - U_n(a_n', \mathbf{a}_{-n})
    \\&= \left[ C_n(a_n, \mathbf{a}_{-n}) - C_n(a_n', \mathbf{a}_{-n})\right] \\
    &+ \sum_{m \neq n} \bigl( C_m(a_m, \mathbf{a}_{-m}) - C_m(a_m, \mathbf{a}_{-m \backslash n}) \bigr)\\
    &- \sum_{m \neq n} \bigl( C_m(a_m, \mathbf{a}_{-m}') - C_m(a_m, \mathbf{a}_{-m \backslash n}') \bigr)
    \\&= \Phi(a_n, \mathbf{a}_{-n}) - \Phi(a_n', \mathbf{a}_{-n}),
\end{aligned}
\end{equation}

Therefore, by iteratively minimizing the overhead $U_n$ of each device own, the system is mathematically guaranteed to reach a pure-strategy Nash equilibrium that aligns with the optimization of the global potential~\cite{chen2025game}.
However, directly calculating the overhead $U_n$ practically incurs prohibitive signaling overhead and delays because the second term in \eqref{utility} requires acquiring the real-time global strategies of all other devices and evaluating their hypothetical cost assuming device $n$ is absent. To overcome this, we propose a novel dual-modality counterfactual inference algorithm to estimate $U_n$ locally.

\subsection{Decentralized Counterfactual Inference Mechanism}
While accurately evaluating the overhead $U_n$ is practically infeasible in highly dynamic LEO networks, ground devices would deduce these unobservable overheads based solely on their local historical observations. However, direct numerical estimations are notoriously unreliable due to data biases within historical data. To overcome this limitation, we develop a fully decentralized, dual-source counterfactual fusion mechanism inspired by~\cite{dou2026dscf}, enabling devices to independently evaluate hypothetical strategies.

\subsubsection{Proxy Counterfactual Dataset Construction}
In highly dynamic LEO networks, a device's historical observational data is heavily influenced by its past offloading policies~\cite{huang2026joint,xu2025edge,li2024computation}. This inherent dependency leads to a spurious correlation between environmental states and action selections. Consequently, direct numerical estimators are prone to conflating the true causal effects of offloading decisions with historical behavioral patterns, yielding severely biased utility evaluations for unexplored strategies. To decouple this relationship and establish a statistically reliable baseline for causal reasoning, we construct a robust proxy counterfactual dataset $\mathcal{D}_{pcf}$. 

Let $\mathcal{D}_{obs}$ denote the historical observational dataset locally maintained by a device, comprising $N$ empirical samples. For the $i$-th sample, we denote its continuous environmental numerical state as $\mathbf{x}_i$ and the executed offloading action as $\mathbf{a}_i$. By defining a concatenation operator $\oplus$, the joint numerical feature representation is formulated as $\mathbf{z}_i = \mathbf{x}_i \oplus \mathbf{a}_i$. Consequently, the observation space is formally defined as:
\begin{equation}
    \mathcal{D}_{obs} = \left\{ \left( \mathbf{z}_i, \mathbf{v}_i^{\text{text}}, U_i \right) \,\middle|\, \mathbf{z}_i = \mathbf{x}_i \oplus \mathbf{a}_i, \forall i \in \{1, \dots, N\} \right\},
\end{equation}
where $\mathbf{v}_i^{\text{text}}$ is the textual state embedding and $U_i$ is the actual recorded overhead.

To simulate an independent counterfactual distribution, we introduce a random permutation operator $\Pi$ over the action sequence to obtain counterfactual interventions $\tilde{\mathbf{a}} = \Pi(\mathbf{a})$. This operation yields a set of synthesized off-support queries $\mathbf{z}_i^{cf} = \mathbf{x}_i \oplus \tilde{\mathbf{a}}_i$. However, purely synthetic queries may fall outside feasible physical constraints. To prevent out-of-distribution generative hallucinations, we project each counterfactual query back to the true physical manifold via an Approximate Nearest Neighbor (ANN) search. The index of the optimal historical anchor $j^*(i)$ is determined by minimizing the squared $L_2$-norm given as
\begin{equation}
    j^*(i) = \mathop{\arg\min}_{k \in \{1, \dots, N\}} \left\| \left( \mathbf{x}_i \oplus \tilde{\mathbf{a}}_i \right) - \mathbf{z}_k \right\|_2^2.
\end{equation}
The exact physical information of the selected anchors forms the proxy sample set. Thus, the proxy counterfactual dataset is mapped as
\begin{equation}
    \mathcal{D}_{pcf} = \bigcup_{i=1}^N \left\{ \left( \mathbf{z}_{j^*(i)}, \mathbf{v}_{j^*(i)}^{\text{text}}, U_{j^*(i)} \right) \right\}.
\end{equation}
This rigorous projection mechanism ensures that the constructed counterfactual distribution successfully breaks the policy bias while remaining strictly anchored to authentic, physically viable generation processes.

\subsubsection{Coarse Retrieval via Numerical Distance}
With the successful construction of $\mathcal{D}_{pcf}$, we establish a comprehensive dual-source knowledge base comprising both the original historical observations and the unbiased counterfactual proxies.
Leveraging this dual-source foundation, during the game evolution, when device $n$ intends to explore a novel, unobserved offloading strategy $a_n'$, the hypothetical decision and the real-time network environment are jointly encoded into a dual-source query $q = (\mathbf{q}^{\text{num}}, \mathbf{q}^{\text{text}})$. To avoid the prohibitive computational overhead of evaluating the entire dataset, we implement a low-complexity numerical filtering mechanism to quickly retrieve highly relevant historical data.

First, we retrieve the top-$K$ nearest neighbors from the observational dataset by calculating the numerical Euclidean distances. This extraction mapping is mathematically defined as:
\begin{equation}
    \mathcal{S}_{obs}(\mathbf{q}^{\text{num}}) = \mathop{\arg\text{Top-}K}_{i \in \mathcal{D}_{obs}} \Big\| \mathbf{q}^{\text{num}} - \mathbf{z}_i \Big\|_2^2.
\end{equation}
Subsequently, an identical retrieval operation is performed on the proxy counterfactual dataset $\mathcal{D}_{pcf}$ to obtain $\mathcal{S}_{pcf}(\mathbf{q}^{\text{num}})$. Finally, the highly relevant candidate set is aggregated by fusing both data sources
\begin{equation}
    \mathcal{S}_{cand} = \mathcal{S}_{obs}(\mathbf{q}^{\text{num}}) \cup \mathcal{S}_{pcf}(\mathbf{q}^{\text{num}}), \quad \text{s.t.} \quad |\mathcal{S}_{cand}| = 2K.
\end{equation}
By extracting this compact candidate set, we effectively narrow down the search space, paving the way for the fine-grained, LLM-empowered semantic Fusion detailed in the next stage.

\begin{algorithm}[t]
\caption{Single-Device Optimization via LLM-assisted Counterfactual Inference and Self-Reflection}
\label{alg:single}
\begin{algorithmic}[1]

\Statex \textbf{Step 1: Initialize}
\State Device $n$, Action space $\mathcal{A}_n$
\State Local observational dataset $\mathcal{D}_{obs}^{(n)}$
\State Pre-trained LLMs $\mathcal{F}_{\text{LLM}}$ and $\mathcal{G}_{\text{LLM}}$
\State Temperature $\tau$, Window capacity $W_{max}$.
\State Initial proxy dataset $\mathcal{D}_{pcf}^{(n)} \leftarrow \emptyset$

\Statex \textbf{Step 2: Proxy Counterfactual dataset Construction}
\For{each historical sample $i$ in $\mathcal{D}_{obs}^{(n)}$}
\State formulate off-support query $\mathbf{z}_i^{cf} = \mathbf{x}_i \oplus \tilde{\mathbf{a}}_i$.
\State Find the optimal anchor $j^* = \mathop{\arg\min}_{k} \|\mathbf{z}_i^{cf} - \mathbf{z}_k\|_2^2$.
\State Append proxy sample $(\mathbf{z}_{j^*}, \mathbf{v}_{j^*}^{\text{text}} \oplus r_{j^*}, U_{j^*})$ to $\mathcal{D}_{pcf}^{(n)}$.
\EndFor

\Statex \textbf{Step 3: LLM-Assisted Counterfactual Optimization}
\For{available action $a \in \mathcal{A}_n$}
\State Construct query $q = (\mathbf{q}^{\text{num}}, \mathbf{q}^{\text{text}})$
\State Construct set $\mathcal{S}_{cand} \leftarrow \mathcal{S}_{obs}(\mathbf{q}^{\text{num}}) \cup \mathcal{S}_{pcf}(\mathbf{q}^{\text{num}})$.
\State \For{each candidate $k \in \mathcal{S}_{cand}$}
\State Compute $s_k = \mathcal{F}_{\text{LLM}}\left(\mathbf{q}^{\text{text}}, \mathbf{v}_k^{\text{text}} \oplus r_k; \mathbf{\Theta}_{\text{LLM}}\right)$.
\EndFor
\State Estimate $\hat{U}_n(a) = \sum_{k \in \mathcal{S}_{cand}} U_k \cdot \frac{\exp(s_k / \tau)}{\sum_{j \in \mathcal{S}_{cand}} \exp(s_j / \tau)}$.
\EndFor
\State Update $a^* \leftarrow \underset{a\in \mathcal{A}_n}{\text{argmin}} \quad \hat{U}_n(a)$.

\Statex \textbf{Step 4: Self-Reflection and Memory Evolution}
\State Execution $a^*$ and obtain $U_n^{\text{true}}$
\State Compute prediction error $\Delta U_n = U_n^{\text{true}} - \hat{U}_n(a^*)$.
\State Generate textual reflection: \\
$r_n = \mathcal{G}_{\text{LLM}}\left( \mathbf{v}_n^{\text{text}}, a^*, \hat{U}_n(a^*), U_n^{\text{true}}, \Delta U_n ; \mathbf{\Theta}_{\text{LLM}} \right)$.
\State Push augmented record $(\mathbf{z}_n, \mathbf{v}_n^{\text{text}} \oplus r_n, U_n^{\text{true}})$ into $\mathcal{D}_{obs}^{(n)}$.
\If{$|\mathcal{D}_{obs}^{(n)}| > W_{max}$}
\State Delete the oldest record from $\mathcal{D}_{obs}^{(n)}$ following FIFO.
\EndIf
\State \Return $a^*$
\end{algorithmic}
\end{algorithm}

\subsection{LLM-Assisted Semantic Fusion Algorithm}
While numerical retrieval successfully identifies structurally similar historical instances, traditional data-driven methods exhibit poor generalization when inferring complex causal externalities (e.g., cascading interference or queuing bottlenecks) inherent to unobserved strategies. To overcome this data dependency and accurately estimate the unobserved overhead, we introduce an LLM-assisted semantic fusion algorithm into the inference pipeline.

\subsubsection{Fine-Grained Semantic Scoring via LLM}
For each numerically filtered candidate sample $k \in \mathcal{S}_{cand}$, we extract its textual state representation $\mathbf{v}_k^{\text{text}}$ alongside its historically generated self-reflection insight $r_k$ (i.e., a natural language critique diagnosing past prediction discrepancies, as will be detailed in Section~\ref{subsubsec:reflection}). Rather than relying on simple geometric distances, we employ an LLM as a sophisticated causal reasoning engine to deeply evaluate the complex semantic alignment between the queried historical context and the unexplored decision. 

Crucially, by reading the historical reflection $r_k$ appended to the state text, the LLM intrinsically avoids repeating previous logical flaws or misjudgments when evaluating similar environments. Let $\mathcal{F}_{\text{LLM}}(\cdot, \cdot ; \mathbf{\Theta}_{\text{LLM}})$ denote the parameterized non-linear reasoning mapping defined by the LLM architecture. The reflection-aware causal similarity score $s_k$ is computed as:
\begin{equation}
    s_k = \mathcal{F}_{\text{LLM}}\left( \mathbf{q}^{\text{text}}, \mathbf{v}_k^{\text{text}} \oplus r_k; \mathbf{\Theta}_{\text{LLM}} \right), \quad \forall k \in \mathcal{S}_{cand}.
\end{equation}
By mapping the augmented textual features into a normalized probability space $s_k \in [0, 1]$, a higher value of $s_k$ quantitatively signifies a stronger causal correspondence and a deeper semantic match.

\subsubsection{Overhead Estimation via Semantic Softmax Fusion}
Finally, to seamlessly blend the unbiased characteristics of the counterfactual proxies with the informational richness of the observational data, we synthesize the expected overhead for the unobserved strategy. 

We apply a softmax normalization, parameterized by a temperature scalar $\tau > 0$, to the LLM-derived semantic scores. The unobserved overhead $\hat{U}_n(a_n' | \mathbf{x}_n)$ is thus robustly derived as the conditionally weighted expectation of the true candidate overheads:
\begin{equation}
    \hat{U}_n(a_n' | \mathbf{x}_n) = \sum_{k \in \mathcal{S}_{cand}} U_k \cdot \frac{\exp\left( s_k / \tau \right)}{\sum_{j \in \mathcal{S}_{cand}} \exp\left( s_j / \tau \right)}.
\end{equation}
By embedding this dual-source, LLM-assisted estimation mechanism back into the exact potential game framework, each ground device can autonomously evaluate its optimal strategy via local inference. This effectively drives the entire LEO-ground network to a stable Nash equilibrium while completely eliminating the dependency on real-time global state interactions.

\subsubsection{LLM-Driven Self-Reflection and Memory Evolution}
\label{subsubsec:reflection}
To enable continuous cognitive adaptation in highly dynamic LEO networks, we further augment the pipeline with an LLM-driven semantic reflection mechanism. Once the inferred optimal strategy $a_n^*$ is physically executed under the specific environmental state, the ground device observes the actual ground-truth overhead $U_n^{\text{true}}$. 

Instead of merely computing a scalar numerical error, the device prompts the LLM to act as a self-critique diagnostic agent~\cite{shinn2023reflexion}. To accurately attribute the root causes of the prediction discrepancy $\Delta U_n = U_n^{\text{true}} - \hat{U}_n$, the LLM requires comprehensive contextual input rather than just numerical metrics. Specifically, the textual environmental state $\mathbf{v}_n^{\text{text}}$, the executed decision $a_n^*$, and the dual utility metrics ($\hat{U}_n, U_n^{\text{true}}$) are jointly encapsulated into a diagnostic prompt. Let $\mathcal{G}_{\text{LLM}}(\cdot ; \mathbf{\Theta}_{\text{LLM}})$ denote the auto-regressive generation mapping of the LLM. The concise textual reflection insight $r_n$ is formally synthesized as:
\begin{equation}
    r_n = \mathcal{G}_{\text{LLM}}\Big( \mathbf{v}_n^{\text{text}}, a_n^*, \hat{U}_n, U_n^{\text{true}}, \Delta U_n ; \mathbf{\Theta}_{\text{LLM}} \Big).
\end{equation}

Driven by this multi-dimensional input, the LLM semantically analyzes hidden physical causalities (e.g., unexpected queuing bottlenecks or underestimated cascading interference) that led to $\Delta U_n$. This derived causal insight $r_n$ is then explicitly appended to the original state text, forming an augmented textual record $\mathbf{v}_n^{\text{text}} \oplus r_n$. 

To prevent memory explosion while continuously evolving the knowledge base, this augmented tuple  $\mathbf{v}_n^{\text{text}} \oplus r_n$ is pushed into $\mathcal{D}_{obs}$ via a First-In-First-Out (FIFO) sliding window, permanently evicting the oldest record upon a new insertion.

Through this continuous auto-calibration mechanism, the local inference baseline dynamically adapts to time-varying satellite topologies. 

To provide a clear roadmap of the proposed scheme, the local counterfactual inference optimization for individual devices is detailed in Algorithm~\ref{alg:single}, while the overarching decentralized potential game framework is summarized in Algorithm~\ref{alg:all}.

\begin{algorithm}[t]
\caption{Decentralized Potential Game Framework for offloading}
\label{alg:all}
\begin{algorithmic}[2]
\Statex \textbf{Step 1: Initialize}
\State Set of ground devices $\mathcal{N}$, 
\State Action spaces $\{\mathcal{A}_n\}_{n \in \mathcal{N}}$
\State Local observational dataset $\{\mathcal{D}_{obs}^{(n)}\}_{n \in \mathcal{N}}$
\State Initial global strategy $\mathbf{a}^{(0)} = (a_1^{(0)}, \dots, a_{|\mathcal{N}|}^{(0)})$ randomly, set $t = 0$.

\Statex \textbf{Step 2: Asynchronous Potential Game Evolution}
\While{not converged}
\State $t \leftarrow t + 1$.
\State \textbf{Asynchronously} select a single device $n \in \mathcal{N}$ 
\State \textit{// Single-device optimization via Algorithm~\ref{alg:single}}
\State $a_n^{(t)} \leftarrow \text{Local Optimization}(n, \mathcal{A}_n, \mathcal{D}_{obs}^{(n)})$.
\State Other devices  strictly maintain strategies: $a_{-n}^{(t)} = a_{-n}^{(t-1)}$.
\State Update global strategy $\mathbf{a}^{(t)} \leftarrow (a_n^{(t)},a_{-n}^{(t-1)})$.
\If{$\mathbf{a}^{(t)} = \mathbf{a}^{(t-1)}$}
\State Global Nash Equilibrium is reached
\State \textbf{Break the While loop}
\EndIf
\EndWhile
\State Set the converged pure-strategy $\mathbf{a}^* \leftarrow \mathbf{a}^{(t)}$.

\end{algorithmic}
\end{algorithm}

\section{Numerical Results}
\label{Numerical}
In this section, we conduct numerical simulations to evaluate our proposed L-COIN offloading framework. We first demonstrate the theoretical convergence and stability of our proposed multi-agent potential game algorithm. Subsequently, we validate the performance superiority of our scheme against state-of-the-art (SOTA) baselines. Finally, we demonstrate the generalization capability of our framework in highly dynamic LEO networks.

\subsection{Simulation Setup and Baseline Algorithms}
Before evaluating the performance of the proposed framework, we first outline the essential simulation setup. In our experiment, we consider the LEO constellation deployed at an altitude $h = 780$~km, consisting of $P = 24$ orbital planes, with $T=15$ satellites per plane. For the ground segment, a total of $N=50$ devices are uniformly and randomly distributed within a specific geographical bounding box (e.g., longitude $118^\circ$E to $122^\circ$E and latitude $28^\circ$N to $34^\circ$N). These devices generate computation tasks with data volumes uniformly distributed in $[1.5, 8.5]$ MB. Additionally, we deploy the pre-trained DeepSeek-Chat LLM model, and the sliding window $W_{max}$ for the local dataset is $50$ per device. All the other parameters are detailed in Table.~\ref{tab1}, unless otherwise specified.

To comprehensively evaluate the superiority of the proposed framework, we compare it with other computation offloading schemes as shown below.
\begin{itemize}
    \item \textbf{CL(Compute Local (CL)):} Each ground device in this method can only compute its task locally.
    \item \textbf{Random:} The task processing approach for each device is random in this method. Specifically, a device is randomly chosen first. Then, the device randomly makes a decision.
    \item \textbf{ICSOC:} Each device in this scheme greedily seeks the best decision to minimize its own cost and updates its strategy iteratively until converged~\cite{lai2019edge}.
    \item \textbf{Ideal:} Each device in this scheme engages in the potential game under the assumption of perfectl global information sharing, serving as the upper bound for game-theoretic approaches~\cite{chen2025game}.
\end{itemize}

\begin{table}[t]
\captionsetup{font={footnotesize}}
\caption{Simulation Parameters}
\label{tab:parameters}
\centering
\begin{tabular}{l l}
\hline  
\hline  
\textbf{Parameter} & \textbf{Value} \\  
\hline 
LEO altitude ($h$) & 780 km  \\ 
Number of orbital planes ($P$) & 24  \\ 
Number of satellites per plane ($T$) & 15 \\ 
Phasing factor ($F$) & 1 \\ 
Orbital inclination ($\theta$) & $86.4^\circ$ \\ 
Carrier frequency ($f_c$) & 220 GHz \\ 
Channel bandwidth ($B$) & 20 MHz \\  
Transmit power ($P_n$) & 23 dBm \\ 
Noise power density ($\sigma^2$) & -114 dBm/MHz \\
Number of ground devices ($N$) & 50 \\
Task data size ($D_n$) & $1.5 \sim 8.5$ MB \\
Softmax temperature ($\tau$) & 0.5 \\
Sliding window capacity ($W_{max}$) & 50 \\
Computation workload per bit ($C_n$) & 500 cycles/bit \\
ground device CPU capacity ($f_n^{loc}$) & 0.5 Gcycles/s \\
Satellite CPU capacity ($f_s$) & 20 Gcycles/s \\
\hline
\hline 
\end{tabular}
\label{tab1}
\end{table}

\begin{figure}[t]
\centerline{\includegraphics[width=0.5\textwidth]{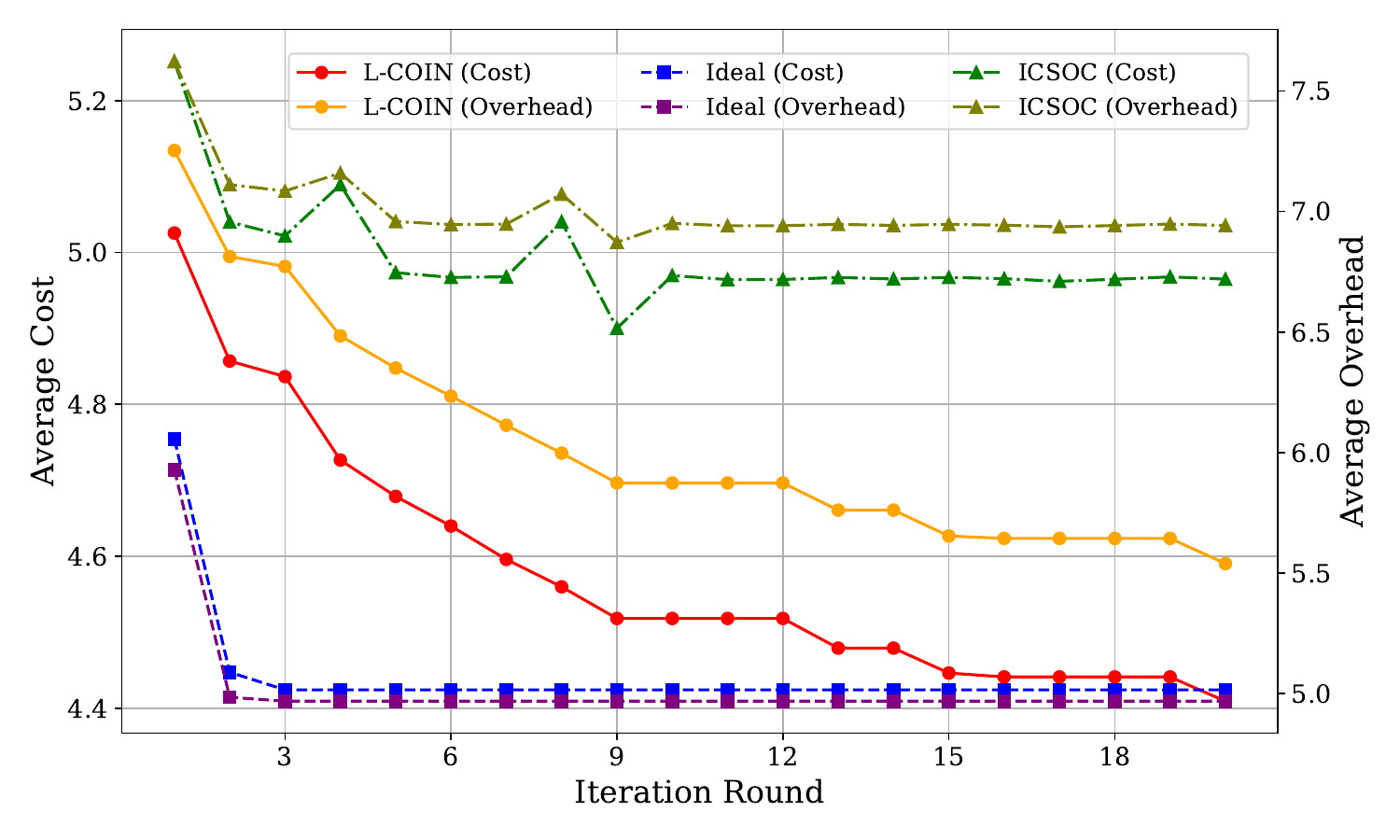}}
\captionsetup{font={footnotesize}}
\caption{Convergence of Average Cost and Overhead under different schemes.}
\label{fig_curve}
\end{figure}

\begin{figure}[t]
\centerline{\includegraphics[width=0.5\textwidth]{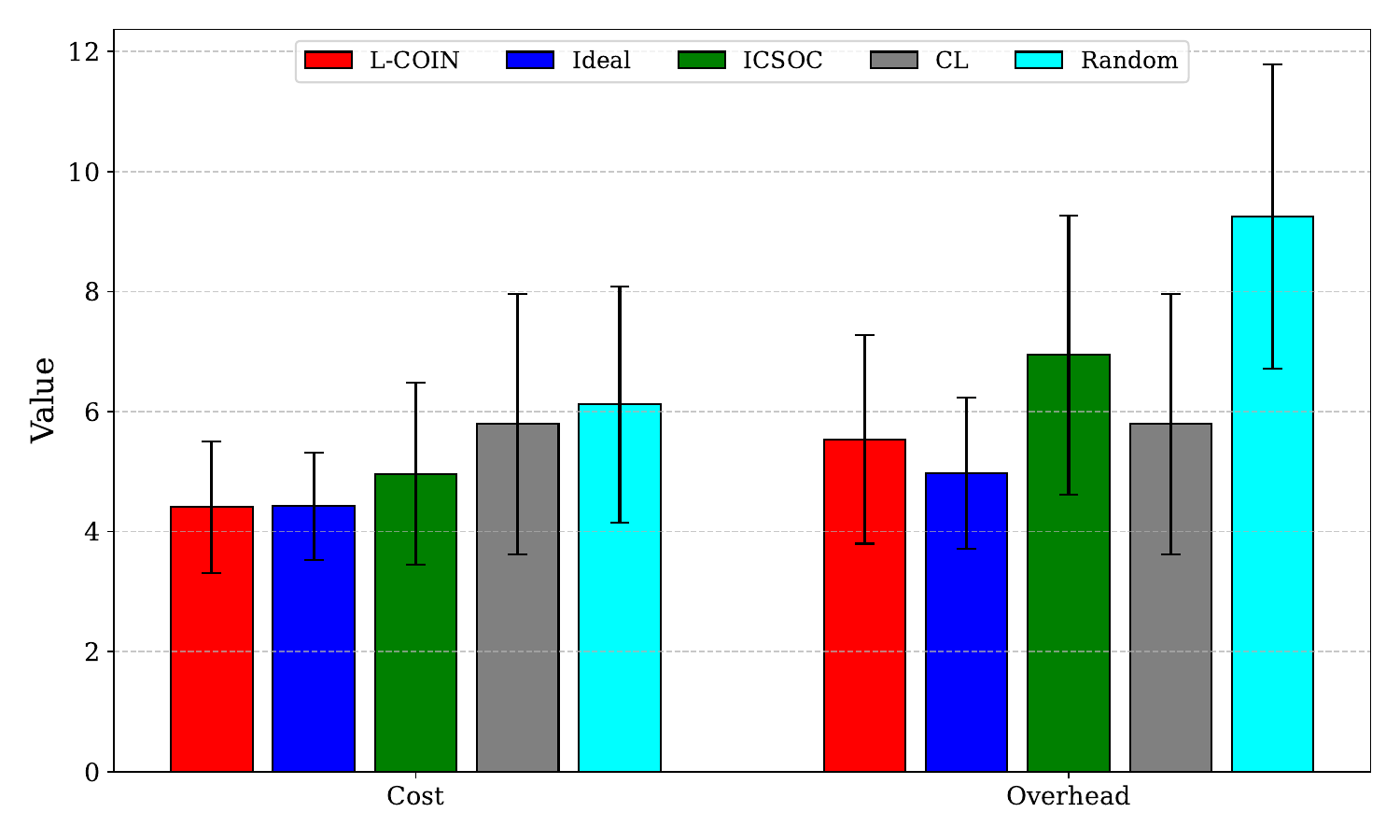}}
\captionsetup{font={footnotesize}}
\caption{Average cost and overhead under different schemes.}
\label{fig_bar}
\end{figure}

\begin{figure}[t]
\centerline{\includegraphics[width=0.5\textwidth]{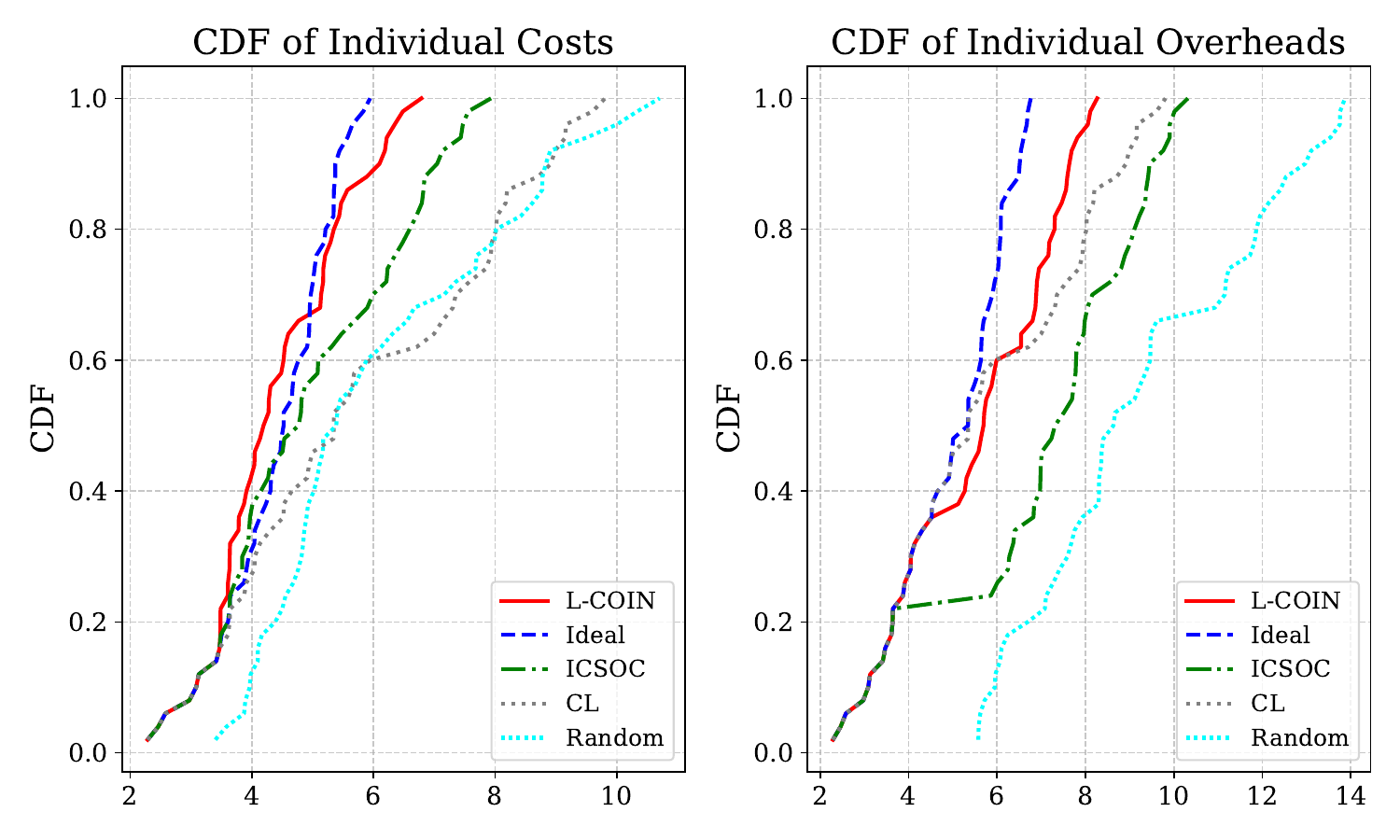}}
\captionsetup{font={footnotesize}}
\caption{CDF of individual device cost of overhead under different schemes.}
\label{fig_cdf}
\end{figure}

\subsection{Convergence and Stability Analysis}
In this subsection, we experimentally verify the theoretical convergence and stability of our proposed L-COIN framework. As observed in Fig.~\ref{fig_curve}, ICSOC experiences occasional slight increases during the iteration process before eventually converging after approximately 10 iterations due to its uncoordinated greedy decisions inevitably cause resource conflicts. In contrast, \textbf{Ideal} demonstrates a strictly monotonic descent and converges to a highly optimal solution in just about 4 iterations since each device has access to the detailed information of all other devices. Notably, our proposed L-COIN framework successfully maintains a strictly monotonic descent in total cost by effectively leveraging LLM-driven counterfactual exploration. Although L-COIN requires approximately 15 iterations to reach full convergence, it ultimately stabilizes at a highly competitive Nash Equilibrium of overhead, which is much lower than ICSOC. This verifies that our proposed L-COIN framework can effectively guide decentralized agents to achieve stable convergence without requiring impractical global information exchange.

\subsection{Performance Comparison and Fairness Analysis}
In this subsection, we evaluate L-COIN against the baseline algorithms, focusing on the average value and distribution of overhead and cost.
First, Fig.~\ref{fig_bar} compares the average offloading cost and overhead, with error bars indicating the standard deviation. CL and Random yield the highest costs around 5.79 to 6.11. ICSOC achieves lower costs of 4.96 but exhibits significant variance due to severe resource congestion caused by uncoordinated myopic decisions. In contrast, \textbf{L-COIN} strictly outperforms all non-ideal baselines by achieving a much lower cost of 4.42, which translates to a significant performance improvement ranging from 10.9\% to 27.7\%. Furthermore, its narrow error bars demonstrate remarkable execution stability, with the achieved cost closely approaching the theoretical lower bound of 4.40 established by the \textbf{Ideal} baseline.

Second, to investigate performance fairness, we plot the Cumulative Distribution Function (CDF) of individual device costs and overhead in Fig.~\ref{fig_cdf}. CL, Random, and ICSOC exhibit a heavy-tail distribution, where nearly $28\%\sim40\%$ of devices suffer from excessively high costs of $6$ due to fierce competition. Conversely, the CDF curve of \textbf{L-COIN} decreases sharply, with only $10\%$ of devices exceeding the cost threshold of $6$. This proves that our framework minimizes the average costs while preventing extreme starvation for individual devices.

\begin{figure}[t]
\centerline{\includegraphics[width=0.5\textwidth]{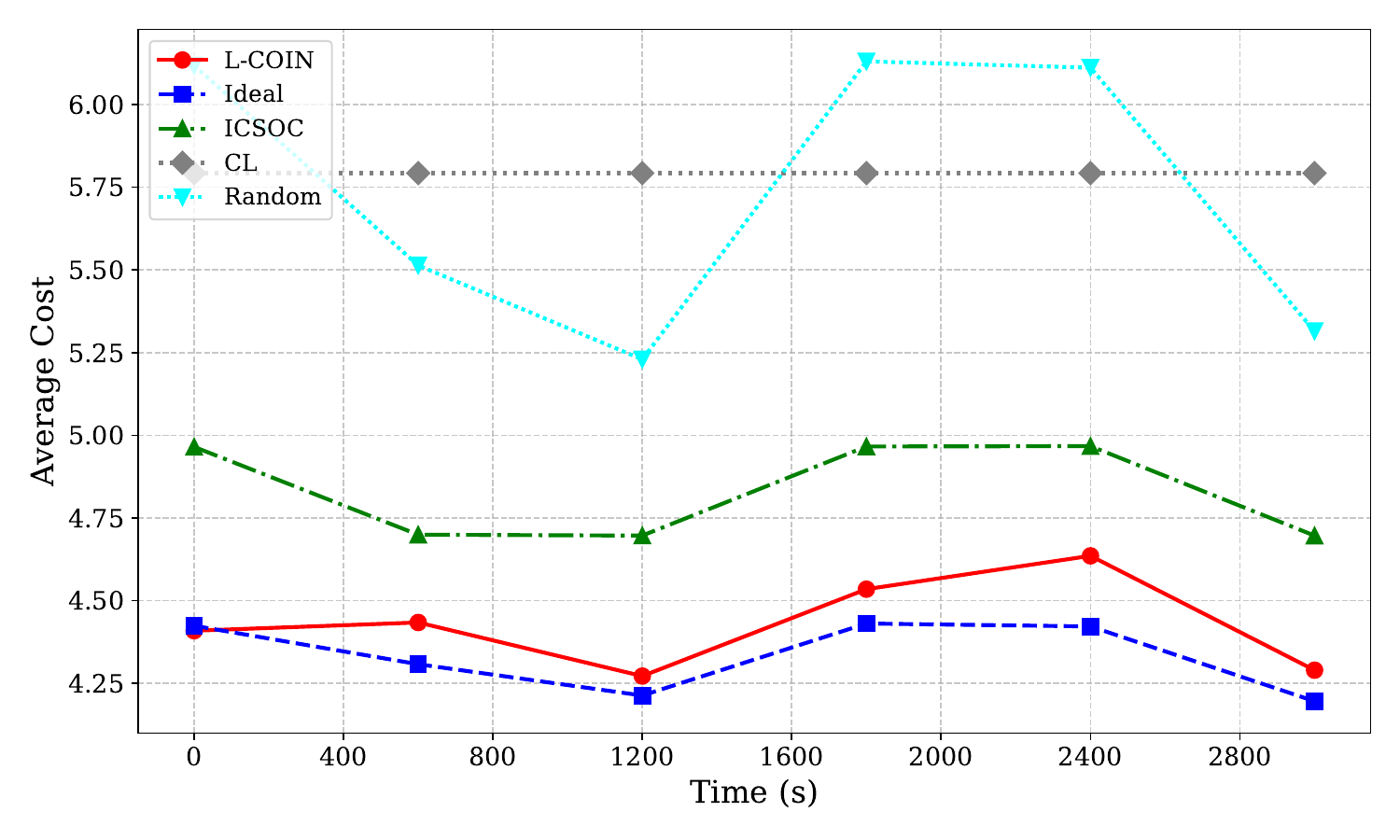}}
\captionsetup{font={footnotesize}}
\caption{Average cost under varying times.}
\label{fig_topology_change}
\end{figure}

\begin{figure}[t]
\centerline{\includegraphics[width=0.5\textwidth]{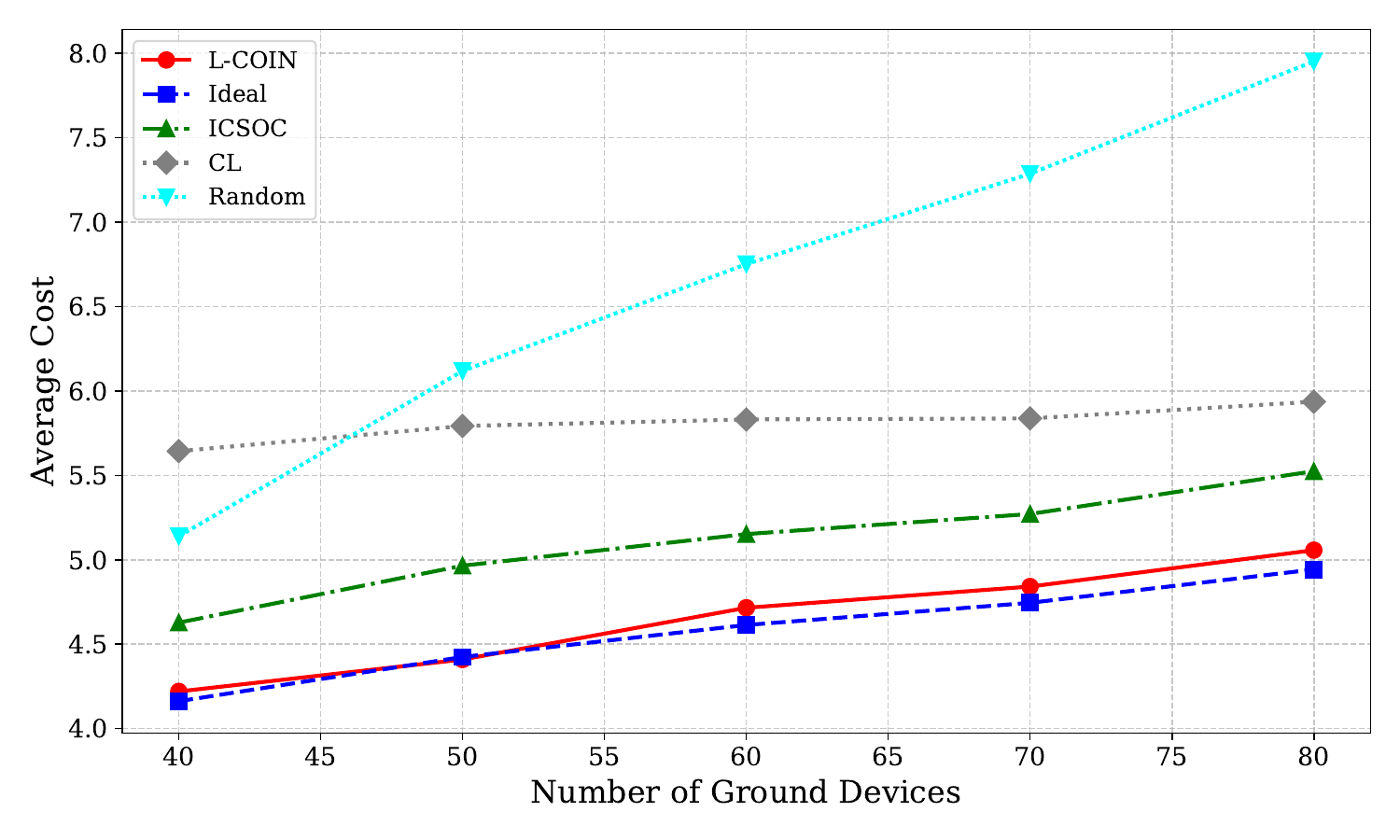}}
\captionsetup{font={footnotesize}}
\caption{Average cost under different numbers of devices and schemes.}
\label{fig_dynamic_devices}
\end{figure}

\subsection{Generalization Capability in Dynamic LEO Networks}
To demonstrate the generalization capability of L-COIN, we evaluate its performance across diverse dynamic scenarios, specifically focusing on varying temporal network topologies and different spatial distributions of ground devices.

Fig.~\ref{fig_topology_change} depicts the average cost across continuous temporal snapshots (from $0$ to $3000$s), reflecting the topological evolution driven by high-speed LEO satellite mobility. While the performance of ICSOC noticeably fluctuates (ranging from $4.70$ to $4.97$) across different topological states due to its rigid greedy nature, \textbf{L-COIN} consistently maintains a stable and low average cost between $4.27$ and $4.64$. By closely tracking the \textbf{Ideal} lower bound ($4.20$$\sim$$4.43$), this confirms that L-COIN can effectively generalize to unseen edge resource distributions without overfitting to specific static topologies.

Furthermore, Fig.~\ref{fig_dynamic_devices} investigates the algorithmic robustness under varying numbers of active ground devices (scaling from $N=40$ to $80$) and their random geographical locations. Changing the device scale and spatial distribution fundamentally shifts the interference and resource competition dynamics, causing a general upward trend in costs. Expectedly, ICSOC struggles to cope with the intensified conflicts, with its cost surging from $4.63$ to $5.52$. Meanwhile, the Random strategy scales poorly, peaking at $7.95$. In contrast, \textbf{L-COIN} maintains a significantly lower cost range of $4.22$ to $5.06$ across all tested device deployments. Notably, even under dense network conditions ($N=80$), \textbf{L-COIN} closely tracks the theoretical lower bound of the \textbf{Ideal} scenario (e.g., $4.94$). By leveraging semantic reflection and the sliding memory window, devices adaptively transfer generalized cognitive insights to infer optimal strategies, proving the strong spatial and temporal generalization capability of the proposed framework.

\section{Conclusion}
\label{Conclusion}
This paper proposes the L-COIN framework, an LLM-empowered distributed computation offloading scheme tailored for highly dynamic LEO-ground edge computing networks. By formulating the resource competition as an exact potential game, the system enables decentralized decision-making among ground devices. To overcome the limitations of incomplete information and myopic greedy behaviors, a pre-trained LLM is integrated as a cognitive reasoning engine, utilizing a sliding memory window to conduct effective counterfactual inference. Numerical results demonstrate that L-COIN successfully eliminates ping-pong oscillations and steadily converges to a superior Nash Equilibrium. In terms of average cost and overhead, L-COIN achieves a low cost of $4.42$, yielding an approximate $25\%$ improvement over non-ideal baselines and closely approaching the \textbf{Ideal} lower bound of $4.40$. Furthermore, it ensures individual fairness and exhibits strong generalization capability across continuous topological shifts and varying device numbers. These findings highlight the critical potential of integrating LLM-driven cognition with game theory to enable robust and efficient decentralized edge computing in future SBINs.

\bibliographystyle{ieeetr}
\bibliography{ref}

\vfill
  
\end{document}